\documentclass[twocolumn,showpacs,amsmath,amssymb,prl,superscriptaddress]{revtex4}

\usepackage{graphicx}
\usepackage{dcolumn}
\usepackage{bm}

\begin{document}

\preprint{PRL}

\title{Coherent optical writing and reading of the exciton spin state in single quantum dots}

\author{Y. Benny}
\author{S. Khatsevich}
\author{Y. Kodriano}
\author{E. Poem}
\author{R. Presman}
\author{D. Galushko}
\affiliation{The Physics Department, Technion - Israel institute of technology, Haifa, 32000, Israel.}
\author{P. M. Petroff}
\affiliation{Materials Department, University of California Santa Barbara, CA, 93106, USA.}
\author{D. Gershoni}
\email{dg@physics.technion.ac.il}
\affiliation{The Physics Department, Technion - Israel institute of technology, Haifa, 32000, Israel.}

\date{\today}

\begin{abstract}
We demonstrate a one to one correspondence between the polarization
state of a light pulse tuned to neutral exciton resonances of single
semiconductor quantum dots and the spin state of the exciton that it
photogenerates. This is accomplished using two variably polarized
and independently tuned picosecond laser pulses. The first ``writes"
the spin state of the resonantly excited exciton. The second is
tuned to biexcitonic resonances, and its absorption is used to
``read" the exciton spin state. The absorption of the second pulse
depends on its polarization relative to the exciton spin direction.
Changes in the exciton spin result in corresponding changes in the
intensity of the photoluminescence from the biexciton lines which we
monitor, obtaining thus a one to one mapping between any point on
the Poincar\'{e} sphere of the light polarization to a point on the
Bloch sphere of the exciton spin.
\end{abstract}

\maketitle

Coherent manipulation of quantum states is a critical step towards applications in quantum information processing.
The atomic-like spectrum of semiconductor quantum dots (QDs) and their compatibility with modern microelectronics make them promising candidates
for forming the building blocks of
these future technologies. In particular, they form an excellent interface between flying photonic-qubits
and anchored matter-spin-qubits.
The coherent properties of spins of confined carriers and pairs of carriers (excitons) in QDs have been demonstrated by various experimental
ways over the years \cite{Kosaka08,Press08,Li03,Boyle08,Flis01,Kamada01,Chen02,Htoon02,Boyle09}.
The fundamental optical excitations of QDs, the exciton and pair of excitons (biexciton),
have been proposed \cite{Biolatti00} and demonstrated~\cite{Li03} as coherent physical realizations
of qubits and quantum logic gates.

\begin{figure}
\includegraphics[width=0.49\textwidth]{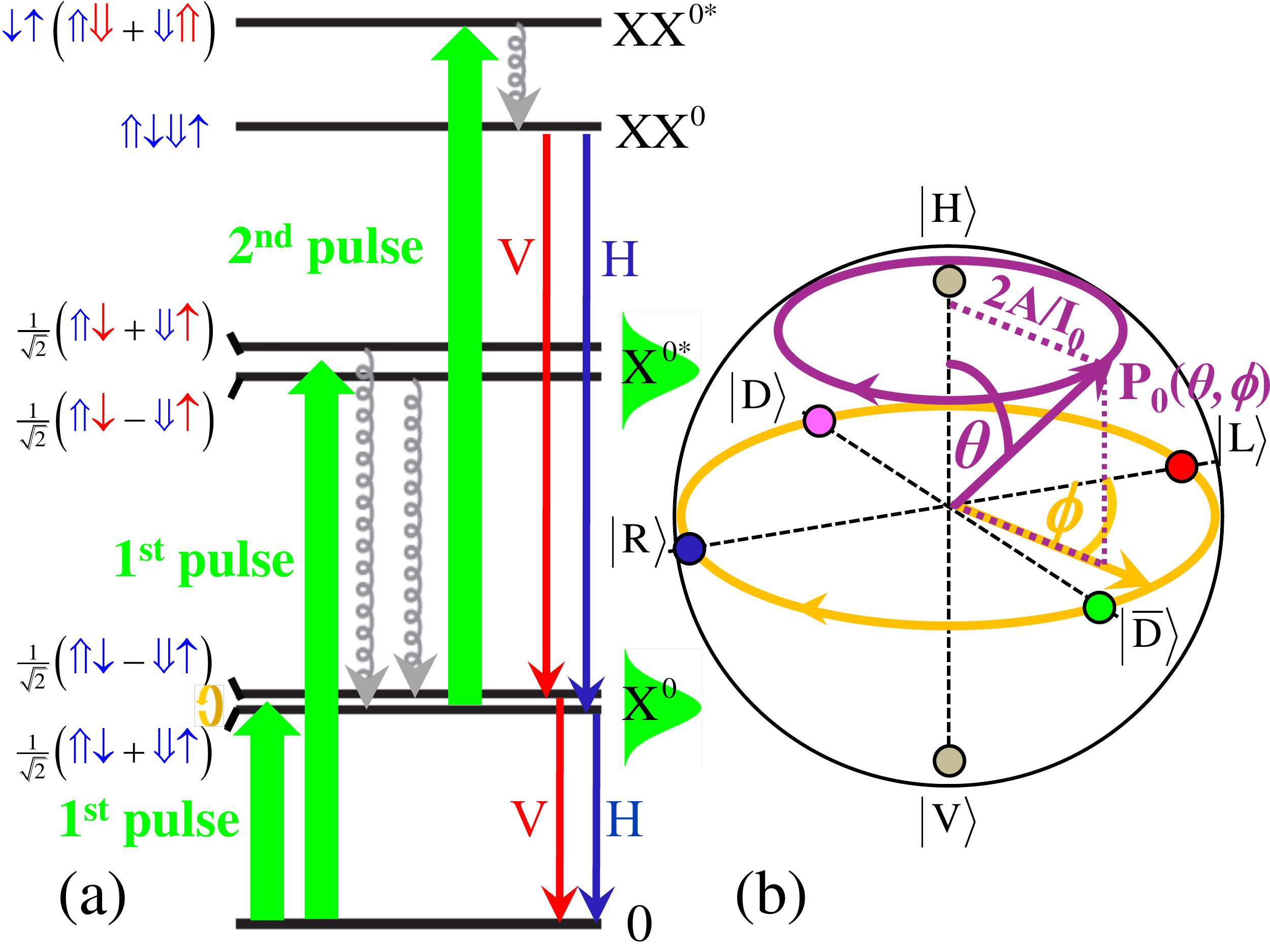}
\caption{\label{fig:EnergyDiagram} (a) Schematic description of the
writing and reading processes. Horizontal lines describe the states'
energies. The spin wavefunctions of these states are depicted to the
left. $\uparrow$ ($\Downarrow$) represents spin up (down) electron
(hole). Blue (red) arrow represents carrier in the ground (excited)
energy level. Green arrows denote resonantly tuned light pulses,
$\rm1^{st}$ short (long) to a ground (excited) exciton and
$\rm2^{nd}$ to an excited biexciton state. Curled lines denote
nonradiative relaxation and a blue (red) line radiative H (V)
polarized recombination. The linewidth of the laser pulses are given
by the curves to the right. (b) A Bloch sphere representation of the
exciton spin state initiated by the polarized laser pulse. The point
$\rm{P_{0}(\theta,\phi)}$ represents an arbitrarily polarized spin
state. $A$ and $\rm{I_0}$, relate to the biexciton PL intensity (see
Fig.\ \ref{fig:OscDelay}(a)).}
\end{figure}

In this work we demonstrate a new method for initializing (``writing'') the spin state of a QD confined exciton in \textit{any} coherent superposition of its eigenstates by a single, resonantly tuned, polarized picosecond light pulse. Likewise,
we show that the spin state of the initialized exciton can be
determined (``readout") using a second, delayed, polarized
picosecond light pulse, resonantly tuned into specific biexcitonic
resonances. Fig.\ \ref{fig:EnergyDiagram}(a) is an energy level
diagram which schematically describes the processes involved in
writing and reading the excitonic spin state. The first polarized
laser pulse is resonantly tuned into either a ground, or an excited
exciton state to photogenerate an exciton. From the excited state,
the exciton rapidly relaxes non-radiatively to its ground state,
while preserving its initial spin (as demonstrated below).

In order to relate the polarization of the light to the spin state
of the photogenerated exciton we note that the angular momentum
projection of right (left) hand circularly polarized light R (L) in
the direction of propagation is 1 ($\rm{-1}$). Upon electron-heavy
hole (e-h) pair generation the electron spin ($1\over{2}$) is
oriented downward (upward) while the heavy-hole spin ($3\over{2}$)
is oriented upward (downward) such that the total angular momentum
projection is conserved. We associate the spin state of such a pair
with the polarization of the light by defining
$\rm|R\rangle=\Uparrow\downarrow$
($\rm|L\rangle=\Downarrow\uparrow)$. Here a spin up (down) electron
is denoted by $\uparrow$ ($\downarrow$) and a spin up (down) heavy
hole is denoted by $\Uparrow$ ($\Downarrow$). With this notation it
is straightforward to show that the correspondence between the
horizontal, vertical, diagonal and cross-diagonal linear
polarizations of the exciting light (H, V, D and $\rm{\bar{D}}$,
respectively) and the spin state of the photogenerated pair are
given by:
$\rm|H\rangle=1/\sqrt{2}(\Uparrow\downarrow+\Downarrow\uparrow)$;
$\rm|V\rangle=-\emph{i}/\sqrt{2}(\Uparrow\downarrow-\Downarrow\uparrow)$;
$\rm|D\rangle=e^{-\emph{i}\pi/4}/\sqrt{2}(\Uparrow\downarrow+
\emph{i}\Downarrow\uparrow)$ and
$\rm|\bar{D}\rangle=e^{\emph{i}\pi/4}/\sqrt{2}(\Uparrow\downarrow-
\emph{i}\Downarrow\uparrow)$. The direction in space of the H
polarization coincides with the direction of the natural major axis
of the QD~\cite{Li03}. These spin states are described on the Bloch
sphere of the exciton spin in Fig.\ \ref{fig:EnergyDiagram}(b). An
arbitrarily elliptically polarized pulse is described by a point on
the surface of the Poincar\'{e} sphere. Such a point can be viewed
as having two components. A component on the equator plane
(containing the L and D directions), deflected by an angle $\phi$
from L, and a component parallel to the rectilinear H-V axis. Thus,
two angles completely define an arbitrary polarization, the angle
$\phi$ and the angle $\theta$ between the polarization and the H-V
axis. In a complete analogy, an arbitrary exciton spin state is
described as a point on the Bloch sphere. The north and south poles
of the Bloch sphere denote the exciton symmetric and antisymmetric
eigenstates, $\rm|H\rangle$ and $\rm|V\rangle$, respectively.
However, due to the anisotropic e-h exchange interaction, these
eigenstates are not degenerate, even in the absence of externally
applied magnetic field~\cite{Bayer02,Li03,Akopian06}

A resonantly tuned H (V) polarized laser pulse photogenerates an
exciton in its symmetric (antisymmetric) spin eigenstate $|H\rangle$
($|V\rangle$). When the pulse resonates with the excited state, the
generated excited exciton relaxes nonradiatively into its ground
state before recombination occurs. The exciton then remains in its
eigenstate until it radiatively recombines. However, since the two
eigenstates are not degenerate, they evolve at different temporal
paces. Therefore, any other coherent superposition of these
eigenstates precesses in time at a frequency given by the difference
between the eigenenergies, divided by the Planck constant. Such
excitation requires, however, a resonant pulse of spectral width
which contains both eigenstates. For example, the orange circle on
the equator of the sphere describes the evolution of an exciton spin
excited by a resonant L polarized pulse excitation. Such a pulse
initiates the two spin eigenstates with equal probabilities. The
initiated spin then precesses clockwise with time, such that the
angle $\phi$ equals $\pi\over{2}$, $\pi$, $3\pi\over{2}$ and $2\pi$
after $1\over{4}$, $1\over{2}$, $3\over{4}$ and $1$ period,
respectively, while the spin state becomes $|\rm{\bar{D}}\rangle$,
$\rm{|R\rangle}$, $\rm{|D\rangle}$ and $\rm{|L\rangle}$ again,
respectively.
The purple circle on the sphere describes precession of an exciton
spin, photogenerated with $\rm{P_0}(\theta, \phi)$ spin by an arbitrarily
polarized pulse.

Delayed by $\Delta\tau$ from the first pulse, a second, polarized
pulse, is then applied. This pulse is tuned to an excited resonance
of the biexciton, to avoid scattered light from the detector. The
probability to photogenerate a biexciton depends on the orientation
of the exciton spin relative to the polarization of the second
pulse, since the biexciton resonance used here contains two
electrons in a singlet state (Fig.\ \ref{fig:EnergyDiagram}).
Therefore, by monitoring the photoluminescence (PL) from the
biexciton doublet as a function of $\Delta\tau$, one obtains direct
information on the evolving exciton spin state.
As schematically described by the point $\rm{P_{0}(\theta,\phi)}$ in
Fig.\ \ref{fig:EnergyDiagram}(b), the spin state of the initiated
exciton can be determined by measuring the phase (thus $\phi$) and
the amplitude (thus $\theta$) of the biexciton signal.

The sample used in this work was grown by molecular-beam epitaxy on
a (001) oriented GaAs substrate. One layer of strain-induced InGaAs
QDs was deposited in the center of a one wavelength
microcavity~\cite{Akopian06,Poem07}. For the measurements the sample
was placed inside a metal tube immersed in liquid Helium,
maintaining temperature of 4.2K. A $\times$60 objective of 0.85
numerical aperture was used to focus the light on the sample surface
and to collect the emitted PL. Two dye lasers, synchronously pumped
at a repetition rate of 76 MHz by the same frequency-doubled
Nd:YVO$\rm_4$ (Spectra Physics-Vanguard$\rm^{TM}$) laser were used
for generating the resonantly tuned light pulses. The duration of
the pulses were 10 ps and their spectral widths about 100 $\rm\mu
eV$. They were continuously tuned using coordinated rotations of two
plate birefringent filters and an etalon. The polarizations of the
pulses were independently adjusted by a polarized beam splitter and
two pairs of computer controlled liquid crystal variable retarders
(LCVRs). The polarization of the emitted PL was analyzed by the same
setup. The delay between the pulses was controlled by a moving
retroreflector. The PL was filtered by a 1-meter monochromator, and
detected by either a silicon avalanche photodetector or a CCD
camera.

\begin{figure}
\includegraphics[width=0.48\textwidth]{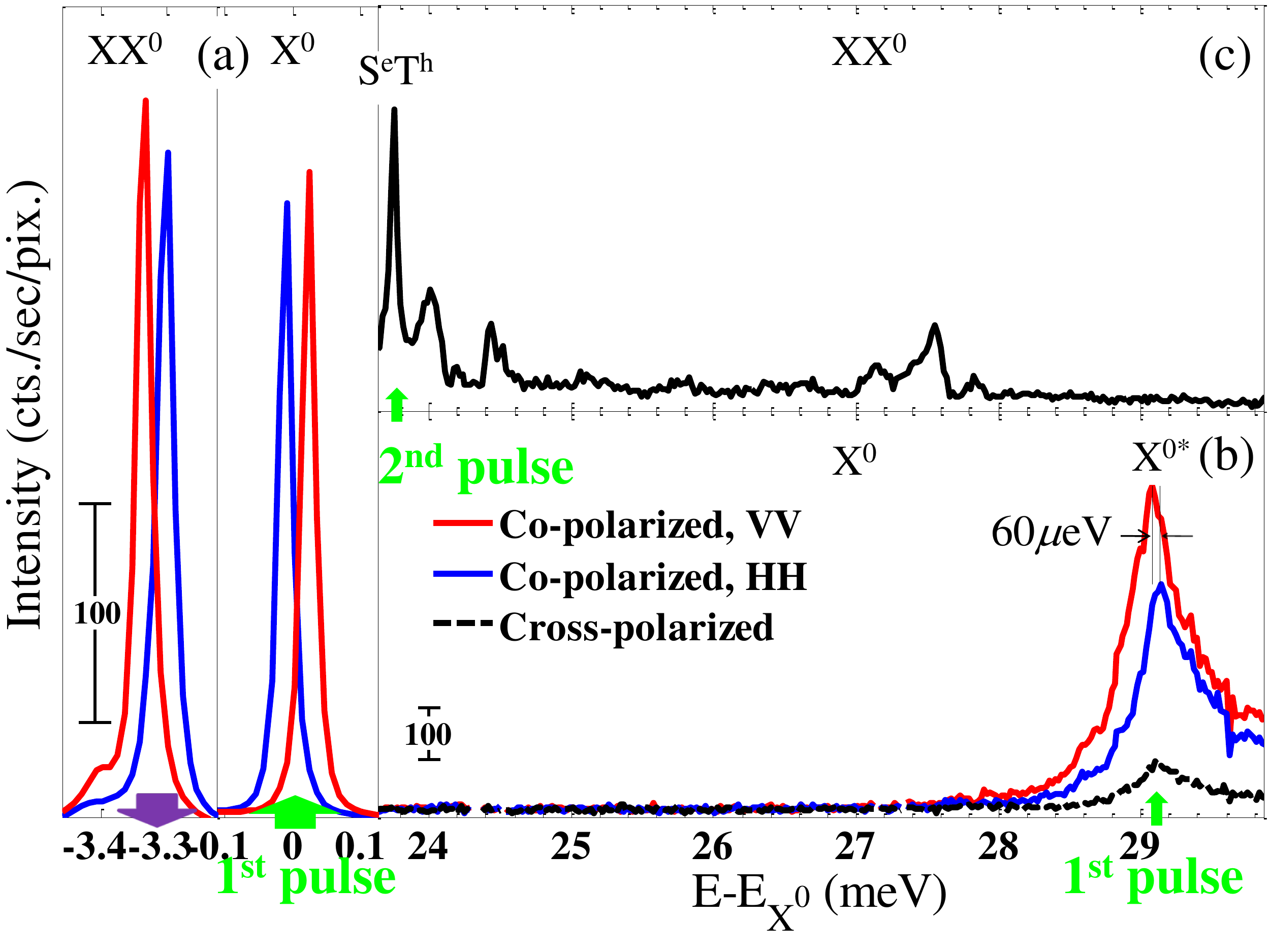}
\caption{\label{fig:PLE_Pulses_PL} (a) V (red) and H (blue)
polarized PL spectra from the resonantly excited QD. (b) [(c)]
Polarization sensitive PLE spectra of the exciton [biexciton]. (c)
was obtained while resonantly exciting the exciton as indicated by
the green vertical arrow in (b). For the readout the second
laser was tuned to the biexciton resonance marked by the vertical
green arrow in (c), while the PL was monitored from the $\rm XX^0$
doublet as marked by the downward purple arrow in (a).}
\end{figure}

Fig.\ \ref{fig:PLE_Pulses_PL}(a) shows polarization-sensitive PL
spectra of a single QD. The neutral exciton ($\rm{X^0}$) and
biexciton ($\rm{XX^0}$) are each indicated with its cross-linearly
polarized doublet. Fig.\ \ref{fig:PLE_Pulses_PL}(b) shows
polarization sensitive PL excitation (PLE) spectra of the exciton
doublet using one laser source. Fig.\
\ref{fig:PLE_Pulses_PL}(c)shows PLE spectrum of the biexciton
doublet using two laser sources. The spectral position of the first
resonant laser, which ``writes" the exciton spin state, was either
tuned to the exciton PL doublet (Fig.\ \ref{fig:PLE_Pulses_PL}(a))
or to its broad resonance (Fig.\ \ref{fig:PLE_Pulses_PL}(b)) as
marked by a green upward arrow.
This strong, linearly polarized resonance, about 29 meV above $\rm
X^0$, originates from the first single hole and second single
electron state (Fig.\ \ref{fig:EnergyDiagram}(a)). Its width is due
to resonant coupling to the first electronic level via one optical
phonon~\cite{Bastard99}. The lifetime of this excited state as
judged by its linewidth ($\rm\sim1$meV) is much shorter than its
precession period as judged by the energy difference between its H
and V co-linearly polarized components ($\rm60\mu$eV - Fig.\
\ref{fig:PLE_Pulses_PL}(b)). The relaxation here, resonantly
mediated by optical phonons, is much faster than that between the
corresponding heavy hole levels, where acoustic phonons mediate
it~\cite{Kodriano10,Poem10}. The initial spin is thus, predominantly
preserved in the relaxation.

The PLE spectrum of the biexciton doublet (Fig.\
\ref{fig:PLE_Pulses_PL}(c))was obtained by scanning the frequency of
the second laser, while the first laser was tuned into excitonic
resonance. The spectral position of the second resonant laser (the
``readout") is indicated by the green vertical arrow on the
biexciton PLE spectrum. Full characterization of the two-photon
biexciton PLE spectrum will be
published elsewhere. 
The particular resonance used here, is due to a ground state
two-electron singlet and a ground and excited state two-heavy holes
triplet [$\rm{S^eT^h}$ - Fig.\ \ref{fig:EnergyDiagram}(a)]. This
resonance differs from the conventional biexciton transition in
which both carrier pairs form singlets in their respective ground
levels. Here, the heavy hole pair form a triplet
\cite{Kodriano10,Poem10}. Therefore, while in the first case the
biexciton is excited through the exciton eigenstates by co-linearly
polarized photon pair, in the latter, the excitation requires a
cross-linearly polarized pair~\cite{Kodriano10}. We use the latter
resonance and not the ground biexciton state, to avoid blinding the
PL detector.

\begin{figure}
\includegraphics[width=0.48\textwidth]{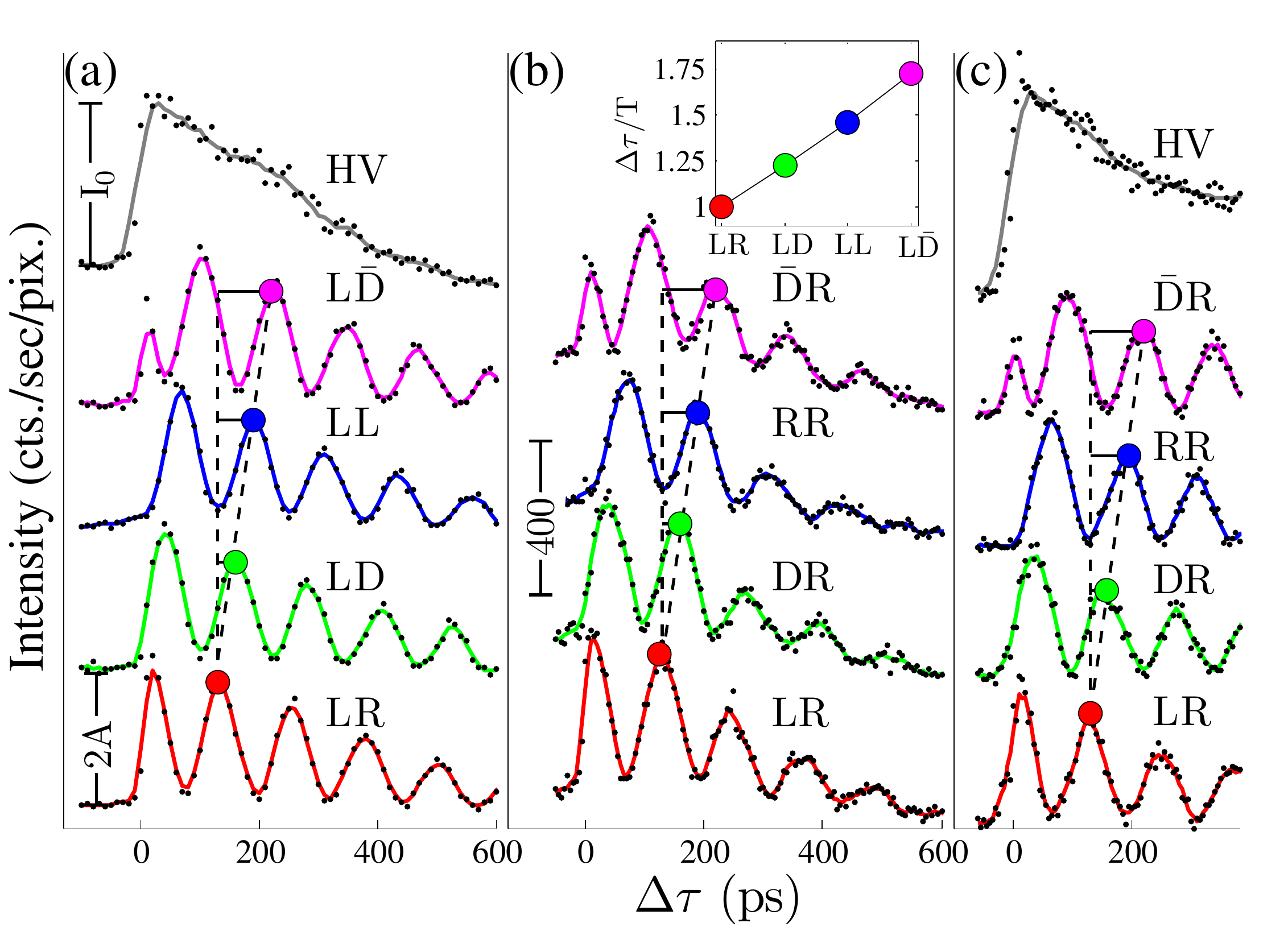}
\caption{\label{fig:OscDelay} (a) Emission intensity of the PL from
the biexciton spectral lines as a function of the delay time between
the pulse into the excited exciton resonance and the pulse into the
biexciton resonance, for various pulse polarizations. The points are
measured data and the solid curves are guides to the eye. The first
(second) letter describes the polarization of the first (second)
pulse. The phase and amplitude of the oscillations are related to
the polarization of the first pulse [see Fig.\
\ref{fig:EnergyDiagram}(b)]. The phase differences between the
various curves are summarized in the inset to (b). (b) Similar to
(a) but for various polarizations of the first laser pulse and fixed
R second pulse. (c) Similar to (b) but here the first pulse is tuned
directly to the ground exciton states.}
\end{figure}

In Fig.\ \ref{fig:OscDelay} we plot the PL emission intensity from
the biexciton doublet as a function of the delay time between the
two laser pulses, for various combinations of the two pulses
polarizations. The first (second) capital letter denotes the
polarization of the first (second) laser pulse. The upper most black
curves in Fig.\ \ref{fig:OscDelay}(a) and (c) present
photogeneration of a biexciton by a cross linearly polarized second
pulse. The biexciton PL signal has maximum immediately after the
first pulse and it decays exponentially as the exciton radiatively
recombines. The rest of the curves in Fig.\ \ref{fig:OscDelay}(a)
show the excitation (``writing") of the exciton by L pulse and
various polarizations of the second pulse which excites the
biexciton (``reads" the exciton spin). In these cases, the first
photon polarization lies on the equator of the Poincar\'{e} sphere,
and it generates a coherent superposition of the two exciton spin
eigenstates with equal probabilities. Since these eigenstates are
energetically separated by 34 $\rm\mu$eV, the exciton spin state
precesses in time along the equator of its Bloch sphere with a
period $\rm{T}=\emph{h}/(34$ $\rm{\mu eV)}=122$ ps. The differences
between the various curves are only in their relative phases. This
dependence is summarized in the inset to Fig.\
\ref{fig:OscDelay}(b). The inset to Fig.\ \ref{fig:OscDelay}
presents the delay times $\Delta\tau$ on which the second maxima are
observed in each one of the four curves in Fig.\
\ref{fig:OscDelay}(a), in units of T. One clearly sees that there is
a constant phase shift of a quarter of a period between the various
polarizations of the second ``readout" pulse.
\begin{figure}
\includegraphics[width=0.48\textwidth]{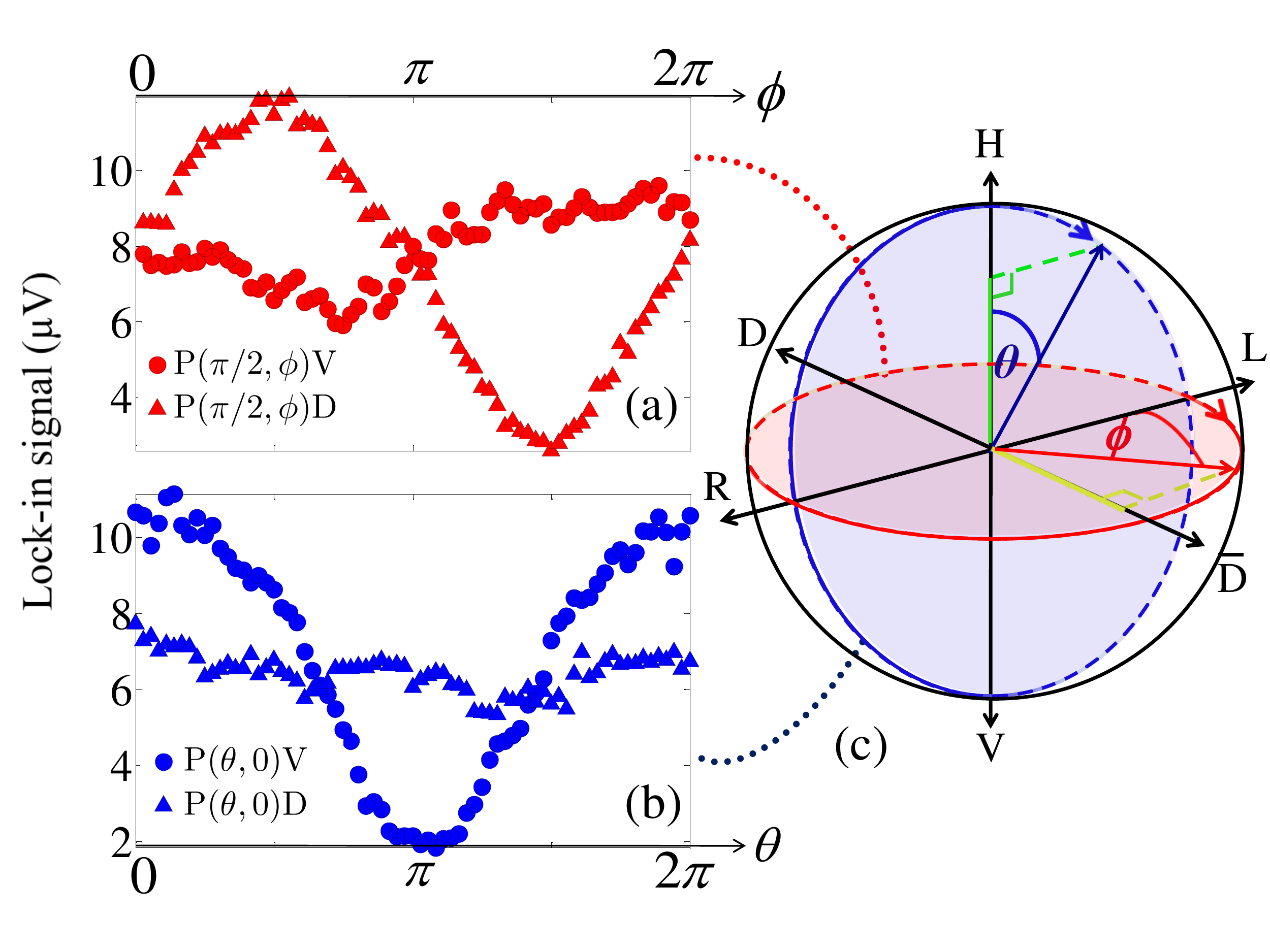}
\caption{\label{fig:OscAngle}  (a) [(b)] biexciton doublet PL
intensity (locked-in to the second laser ) as a function of the
polarization angle $\phi$ [$\theta$] as defined in (c), for D
(triangles) and for V (circles) polarized readout pulses.
$\Delta\tau$ between the two pulses was T=122 ps.}
\end{figure}
The probability to photogenerate a biexciton by a second pulse
depends on the pulse polarization with respect to the exciton spin
polarization. Therefore, the second pulse effectively ``projects"
the exciton spin state onto the complementary direction. This means
that a R (L) pulse projects the spin onto the $\rm|L\rangle$
($\rm|R\rangle$) state and a $\rm{\bar{D}}$ (D) pulse projects the
spin onto the $\rm|D\rangle$ ($\rm|\rm{\bar{D}}\rangle$) state.
Similarly, a H (V) pulse projects the spin onto the $\rm|V\rangle$
($\rm|H\rangle$) state. Thus, the state of any polarized exciton
spin can be determined by the polarization of the initial pulse and
a second projective light pulse. Similar behavior is observed when
one varies the polarization of the writing pulse while keeping fixed
the polarization of the readout pulse, as we show in Fig.\
\ref{fig:OscDelay}(b) and (c). Fig.\ \ref{fig:OscDelay}(c) clearly
shows that there is no difference in the exciton spin evolution when
it is excited into the ground or into the excited state. This proves
that the non-radiative relaxation is mostly spin conserving. Our
experimental measurements demonstrate unambiguously that a
resonantly tuned H, V, D, $\rm{\bar{D}}$, R or L picosecond pulse
photogenerates (writes) an exciton with initial spin state
$\rm|H\rangle$, $\rm|V\rangle$, $\rm|D\rangle$,
$\rm|\bar{D}\rangle$, $\rm|R\rangle$, or $\rm|L\rangle$,
respectively.

In Fig.\ \ref{fig:OscAngle} we present a set of measurements in
which an arbitrary initial positioning of the exciton spin on its
Bloch sphere is demonstrated. The figure describes writing of the
exciton spin state by continuous variation of the polarization of
the first pulse along a given circle on the Poincar\'{e} sphere
while leaving the temporal delay between the pulses fixed at
$\rm\Delta\tau=T$. The polarization of the readout pulse is left
fixed, either in a normal direction to the plane of variation of the
first pulse or in that plane. In the first case, one expects the
biexciton PL signal to remain constant since the spin projection on
the probe direction is constant, independent of the in-plane angle.
In the second case, however, the spin projection on the probe
direction is expected to undergo maximal periodic oscillations
resulting in the largest amplitude of oscillations of the signal of
the biexciton. In one set of measurements [Fig.\
\ref{fig:OscAngle}(a)] the polarization angle is varied about the
V-H axis ($\phi$ in the L-D plane) and in the other set [Fig.\
\ref{fig:OscAngle}(b)] the angle is varied about the
D-$\rm{\bar{D}}$ axis ($\theta$ in the H-L plane). The readout in
both cases is performed with D pulse (triangles) and with V pulse
(circles). The observed oscillations can be described as change in
the magnitude of the projection of the Bloch vector along the
direction of the polarization of the readout pulse. The points on
the equator, the circle that is defined by varying $\phi$, while
leaving $\theta=\pi/2$ have the same projection (equal to 0) on the
V direction. Therefore, almost no oscillations are observed in this
case. The projection on the D direction, however, undergoes maximal
variations with the angle $\phi$. Indeed, large periodic
oscillations in the signal are observed. Clear maxima (minima) in
the intensity are obtained when the Bloch vector of the exciton is
antiparallel (parallel) to that of the probe.  As expected for this
particular biexcitonic resonance, maximum absorption is obtained for
cross-linear polarizations \cite{Kodriano10}. In a complementary set
of measurement [Fig.\ \ref{fig:OscAngle}(b)] the opposite behavior
is observed. Here the angle $\theta$ is continuously varied, while
$\phi= 0$. Now, maximal oscillations occur for the V readout pulse
and diminishing oscillations for the D readout pulse. The small
oscillations observed in the latter are probably due to small
inaccuracies in the calibration of the LCVRs and in the alignment of
the setup axes relative to those of the QD. Fig.\ \ref{fig:OscAngle}
demonstrates that the spin state of the exciton can be prepared at
any point on the Bloch Sphere [Fig.\ \ref{fig:OscAngle}(c)], in
correspondence to the elliptic polarization of the writing light
pulse on its Poincar\'{e} sphere.

In summary, we establish clear correspondence between the
polarization of a light pulse tuned to excited or ground excitonic
resonances, and the initial spin state of the photogenerated
exciton.
We directly map the polarization of the light pulse, as represented
by a point on the Poincar\'{e} sphere, to exciton spin state as
represented by a point on the Bloch sphere. For this we use a
second, delayed polarized pulse tuned to particular -
electronic-singlet, biexcitonic resonance. The second pulse projects
the excitonic spin state onto a predetermined direction, providing
thus a way for reading the exciton spin.

\begin{acknowledgments}
The support of the US-Israel binational science foundation (BSF),
the Israeli science foundation (ISF), the ministry of science and
technology (MOST) and that of the Technion's RBNI are gratefully
acknowledged.
\end{acknowledgments}


\end{document}